\documentstyle[epsfig]{article}                                            
\begin{document}

\begin{center}
{\bf INSTITUT~F\"{U}R~KERNPHYSIK,~UNIVERSIT\"{A}T~FRANKFURT}\\
D - 60486 Frankfurt, August--Euler--Strasse 6, Germany
\end{center}

\hfill IKF--HENPG/5--97

\vspace{0.5cm}
\begin{center}
   {\large \bf On Event--by--Event Fluctuations in Nuclear Collisions } \\[2mm]
   Marek Ga\'zdzicki$^1$, Andrei Leonidov$^2$ and Gunther Roland$^1$ \\[5mm]
   {\small \it  $^1$Institut f\"ur Kernphysik, Universit\"at Frankfurt \\
   August--Euler Str. 6, D--60486 Frankfurt/M, Germany \\
   $^2$P.N.~Lebedev Physics Institute, 117924 Leninsky pr. 53\\
   Moscow, Russia\\[8mm] }
\end{center}

\begin{abstract}\noindent

\end{abstract}

We demonstrate that a new type of analysis in heavy--ion collisions,
based on an event--by--event analysis of the transverse momentum
distribution, allows us to obtain information on
secondary interactions and collective behaviour that is not available from
the inclusive spectra.
Using a random walk model as a simple phenomenological description of
initial state scattering in collisions with heavy nuclei, we show
that the event--by--event measurement allows a quantitative determination
of this effect, well within the resolution achievable with the new
generation of large acceptance hadron spectrometers.
The preliminary data of the NA49 collaboration on transverse momentum
fluctuations indicate qualitatively different behaviour than that
obtained within the random walk model.
The results are discussed in relation to the thermodynamic and
hydrodynamic description of nuclear collisions.

\section{Introduction}

The ultimate goal of present heavy ion experiments at the CERN SPS
and similar studies foreseen at future
collider facilities is the production and characterisation of an
extended volume of deconfined quarks and gluons, the quark gluon plasma
(QGP). The QGP has been predicted as the equilibrium state of strongly
interacting matter at sufficiently high temperature and density \cite{qgp}.
A variety of possible signatures for the transient existence
of a deconfined phase of matter in the course of nucleus--nucleus
(A+A) collisions
has been proposed theoretically and studied experimentally (see
\cite{qm96} for a recent review).\\
An important aspect of these studies is answering the question
to which extent  and at which stage of the interaction
thermal and chemical equilibration
is reached in nuclear collisions.
Recent experimental results show a  significant change of the
shapes of particle spectra and relative particle yields
when going from nucleon--nucleon (N+N) interactions to
central collisions
of heavy nuclei \cite{Jo:96}.
These observations find a natural interpretation within statistical
(equilibrium) models assuming thermal and chemical equilibration of
 produced matter
(hadron gas \cite{Le:90,Be:97} or QGP \cite{Ga:95}).
However there are numerous attempts to describe the data without
invoking equilibration and assuming that the observed effects
are only due to initial, non--equilibrium, scatterings.
In the case of inclusive observables equilibrium and non--equilibrium
approaches lead to a qualitatively similar behaviour and
therefore they are difficult to distinguish, if the absence
of the caluclable theory in the domain of soft QCD processes
is taken into account.

It was realized \cite{GaMo92} that
a study of event--by--event fluctuations may serve as a tool to measure
equilibartion in nuclear collisions and therefore allow to distinguish
between equilibrium and non--equilibrium models.
In this paper we follow this approach and consider transverse
momentum fluctuations in non--equilibrium, initial state scattering
models and in equilibrium, hydrodynamical models.

It is  experimentally established \cite{Jo:96} that the inverse slope of transverse
momentum distribution  increases when going from p+p interactions to
central Pb+Pb collisions. The strongest effect is observed for heavy
hadrons.
The  equilibrium models relate this observation
to the presence of transverse hydrodynamical expansion
of the matter \cite{Le:90}. The size of this effect
increases with the size of the colliding nuclei and is most
pronounced in the transverse momentum spectra of the
heaviest particles.
Such models, which apart from the choice of an appropriate flow
profile, can be characterized by two free parameters, the temperature
at freeze--out and a flow velocity parameter, have been shown to
fit the  measured inclusive spectra \cite{Ro:97, Al:97}. \\
However,
it has been demonstrated that the same spectra
can also be fitted by a
different class of models, which do not require the existence of equilibration
\cite{leonidov, kapusta}.
These models are triggered by  the observation of a broadening of transverse
momentum distributions already in proton--nucleus (p+A) interactions ('Cronin--effect'),
which is commonly attributed to  multiple scattering of the incident nucleon
and its remnants in the target nucleus \cite{Cr:73}.
Several authors have parametrized the initial state scattering, which
effectively translates
part of the longitudinal momentum of the incident nucleon into transverse
momentum, in terms of a 'random walk' in a transverse kinematical variable
such as a transverse rapidity \cite{leonidov} or transverse velocity
 \cite{kapusta}.
 The parameters of the random walk can be fixed by a fit to p+A data
 \cite{leonidov}, leading to a fit of
A+A spectra that is essentially parameter free, if the inverse
slopes from N+N interactions are used as starting values. Keeping
this difference in mind, the initial state scattering or random walk models
provide a description of the data that is of similar quality as that of the
hydrodynamical models. A study of inclusive particle
spectra alone is therefore clearly not enough to understand whether
the produced particles originated from a source having a sufficient
amount of secondary interactions to develop collective behaviour such as
hydrodynamical transverse flow.
These two approaches can however be distinguished by analysis
of two pion correlations \cite{To:97} and
event--by--event fluctuations as shown in this paper.

The paper is organized as follows.
In Section 2 we recall the method introduced in Ref. \cite{GaMo92}
and used further in the paper to study event--by--event fluctuations.
The fluctuations in an initial state scattering model are calculated
in Section 3.
The discussion of fluctuations in thermal models is given in Section 4.
Summary and conclusions close the paper.

\section{The Measure of Event-by-Event Fluctuations}

Event--by--event fluctuations in nuclear collisions are usually dominated by
the trivial variation in impact parameter from event to event and the purely
statistical variation of the measured quantities. A statistical method that
allows us to remove  these trivial contributions and to  determine the
dynamical event--by--event fluctuations has been proposed in \cite{GaMo92}.
 Following this reference we define for every particle $i$:
\begin{eqnarray*}
z_i = p_{T_i} - \overline{p_T},
\end{eqnarray*}
where $\overline{p_T}$ is the mean transverse momentum of accepted particles
averaged over all events (the inclusive mean).
Using $z_i$ we calculate for every
event
\begin{eqnarray*}
Z = \sum_{i=1}^N z_i,
\end{eqnarray*}
where $N$ is the number of accepted particles in the event. With this definition
we obtain the following measure to characterise the degree of fluctuation
in the $p_T$ distribution from event to event:
\begin{eqnarray}
\Phi_{p_T}  = \sqrt{\frac{\langle Z^2 \rangle}{\langle N \rangle}} -
\sqrt{\; \overline{z^2}},
\label{Phi}
\end{eqnarray}
where $\langle N \rangle$ and $\langle Z^2 \rangle$ are averages over all
events
and the second term in the r.h.s. is the square root of the second moment
of the inclusive transverse momentum distribution.\\
The physical motivation for studying $\Phi_{p_T}$ was given in \cite{GaMo92}:
experimental data on N+N interactions show
that particles in these collisions are not produced independently \cite{Ka77}.
One observes large scale correlations that lead to, e.g.,\ a correlation
between the event multiplicity and the average $p_T$ of the particles.
This property can be used to probe the dynamics of nuclear collisions
by
measuring to which degree the correlations present in N+N interactions are
changed when going to p+A and A+A collisions.\\
For this purpose, $\Phi_{p_T}$ as a measure of
fluctuations has two important properties. For a large
system (i.e.\ a A+A collision) that is a superposition of
many independent elementary systems (i.e.\ N+N interactions),
$\Phi_{p_T}$ has a constant value that is identical to that
of the elementary system.
In other words if the central  Pb+Pb collisions were a incoherent superposition
of N+N interactions, the value of $\Phi_{p_T}$ would
remain constant, independent of the number of
superimposed elementary interactions
in a single event and its distribution in the studied
sample of the events.
If on the other hand the large system consists of particles
that have been emitted independently, $\Phi_{p_T}$ assumes
a value of zero.
Thus $\Phi_{p_T}$  provides us with a scale
characterising the fluctuations in nuclear collisions
relative to elementary interactions at the same
energy.\\
One should expect that $\Phi_{p_T}$ is sensitive to the initial
state scattering effects considered in \cite{leonidov} and \cite{kapusta}
through additional fluctuations in the transverse momenta of effective
N+N sources superimposed in these models. As the amount of rescattering
is directly related to the size of the colliding nuclei, here one expects
a smooth dependence of $\Phi_{p_T}$ on the nuclear atomic number.
Finally, if the underlying physical mechanism for A+A collisions
corresponds to some different type of correlations
between the (average) transverse momentum and multiplicity
 such as discussed, e.g., in the context of hydrodynamical models
\cite{Sh, VH, Gyu} and further in \cite{VR}, then one could
expect an abrupt change in $\Phi_{p_T}$,
when crossing a boundary in the size of colliding nuclei and/or
collision energy,
where the
new physical mechanisms beyond those in elementary interactions are
already at work.

\section{Event--by--Event Fluctuations  \\ in Initial State Scattering Models}

The motivation for applying this method to test 
the initial state scattering models is that particles in these models are
not produced independently. They are emitted from sources that,
after an initial scattering, are moving in transverse direction.
This transverse source movement is superimposed on the
particle spectrum emitted by the individual particle source,
where apart from the overall boost in transverse rapidity
 each N+N--like source is characterized by the same correlation of
transverse momentum vs multiplicity as for N+N interactions. Thus
the pattern of particle emission changes. A new source of fluctuations
is added (initial state scattering) and we expect that the fluctuation measure
$\Phi_{p_T}$ from Eq.\ (\ref{Phi}) will be sensitive to its presence.

To get an estimate of the size of this effect we use as
reference a simple parametrization of p+p data at
$\sqrt{s} \approx 20$ GeV/c as proposed in Ref. \cite{GaMo92}.
We restrict ourselves to a study of negativly charged hadrons, which
for this energy are mostly ($>90\%$) pions.
The transverse momentum, $p_T$, probability distribution for
events with negatively charged hadron multiplicity $n$ is
parametrized as:
\begin{equation}
h_{n}(p_T) = C_n p_T e^{- m_T/T(n) },
\end{equation}
where $C_n$ is a normalization factor, $m_T$ is a transverse mass
($m_T = \sqrt{(p_T^2 + m^2)}$, where m is the pion mass) and
$T(n)$ is the inverse slope parameter.
The $T(n)$ dependence on multiplicity is taken to be:
\begin{equation}
T(n) = T_0 + (n-1) \frac {\Delta T} {\Delta n},
\end{equation}
where  parameters $T_0$ and $\Delta T/ \Delta n$ are taken to be
173 MeV and -6 MeV in accordance to p+p data at $\sqrt{s} \approx 20$ GeV/c
\cite{Ka77}.
The multiplicity distribution is calculated using the parametrization
given in Ref. \cite{Ga:91}.

We use a Gaussian rapidity distribution  to
describe the longitudinal momentum distribution of the produced hadrons.
To check the sensitivity of our results, we use two extreme values for the width
of the rapidity distribution, $\sigma_y = 1.6$ which corresponds to the
value observed in p+p, and $\sigma_y = 0.8$, which is closer to
isotropic particle emission in the source rest frame.

Starting from this type of elementary particle sources, we then describe
a collision of heavier systems following the model presented in
\cite{leonidov}.
In this initial state scattering model one takes into account
the rotation of the collisions axis of projectile nucleons that have already
undergone a collision with a target nucleon. The effect of multiple collisions
is parametrized by giving the sources of produced particles ('clusters'
or 'fireballs') a Gaussian distribution in transverse rapidity $\rho$ \cite{leonidov}:
\begin{equation}
f_{AB}(\rho) = [ \frac{4}{\pi\delta^2_{AB}}]^{1/2} e^{(-\rho^2/\delta^2_{AB})},
\end{equation}
using $\delta_{AB} = (N_A+N_B -2)\delta^2$.
Here $N_A$ and $N_B$ are the numbers of interacting nucleons from
target
and projectile nuclei
and $\delta$ is the average shift in transverse rapidity per collision.

As has been shown in \cite{leonidov}, this model is able to quantatively describe
the $p_T$--broadening observed in collisions of heavy nuclei, using
a value for $\delta$ of 0.22 that was fixed by fitting spectra
from p+W collisions.

The result of our simulation based on the model described above are
summarized in Fig.~1. It shows the evolution of $\Phi_{p_T}$
with the width of the source transverse rapidity distribution (i.e.\
increasing system size).
 Only particles with $\mid y_{CMS} \mid < 3$ and $p_T < 3$~GeV/c
were used for the calculation of $\Phi_{p_T}$.
\begin{figure}[h]
\mbox{\epsfig{file=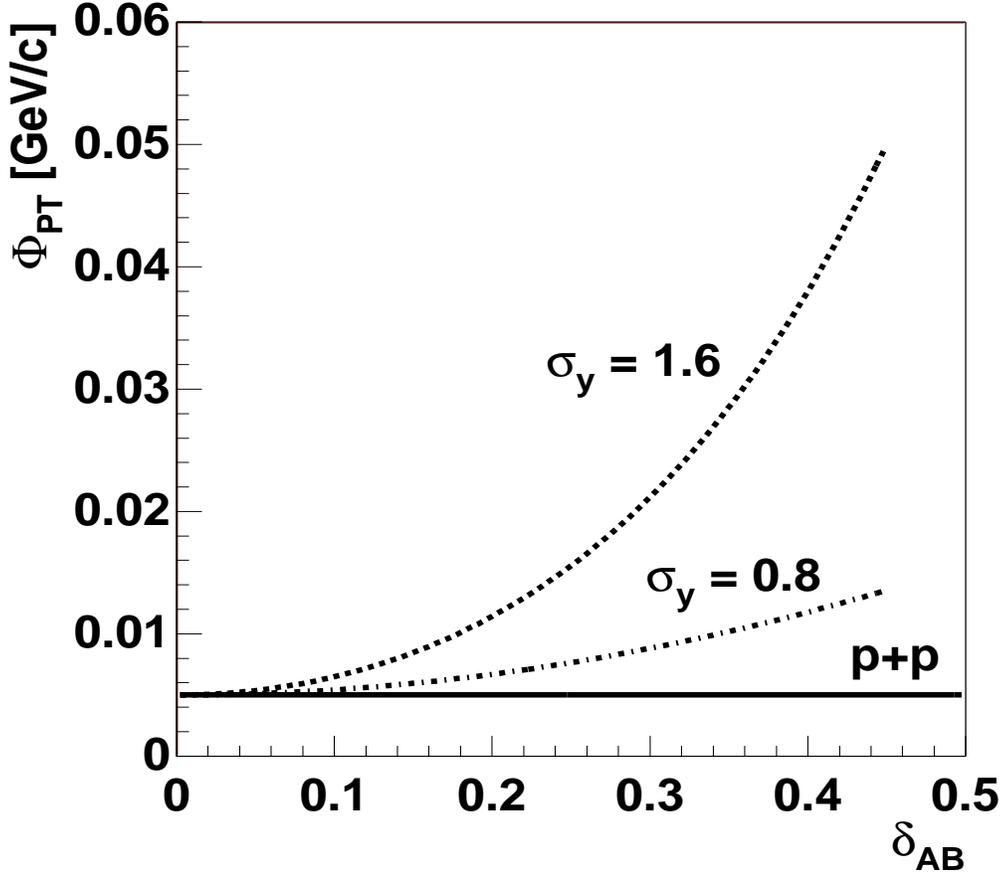,width=15cm,height=13cm}}
\caption{Dynamical $p_T$ fluctuation measure $\Phi_{p_T}$ for the random walk
model with $\sigma_Y = 1.6$  and $\sigma_y = 0.8$
as a function of $\delta_{AB}$.
The soild line indicates the results without initial state scattering
(a linear superposition  of p+p interactions).}
\end{figure}
The solid line in Fig. 1 indicates the reference value of
 $\Phi_{p_T} \approx 4.5$~MeV/c, which we obtain using our parametrization
of p+p data.
 This value is the result of the correlation between average transverse
 momentum and mean multiplicity in the parametrization.\\
For both values of $\sigma_y$ we observe a strong enhancement of $\Phi_{p_T}$
 above this reference value for the random walk model, reaching 10~MeV/c
 for $\sigma_y = 0.8$ and 50~MeV/c for $\sigma_y = 1.6$ at a value of
 $\delta_{AB} = 0.44$, which corresponds to central Pb+Pb collisions.\\
 The  conclusion we draw is that the introduction
of the random walk source movement does not tend to wash out the correlations
present in p+p interactions.
On the contrary the correlation introduced by the transverse source
velocity leads to an amplification of large scale fluctuations in
transverse momentum even by a factor of up to 10.

The preliminary results of the NA49 Collaboration \cite{Rol:97}
show a qualitatively different behaviour. The $\Phi_{p_T}$ value
obtained for central Pb+Pb collisions at 158 A$\cdot$GeV
is significantly smaller than the corresponding value estimated for p+p
interactions.
This interesting result indicates that the non--equilibrium
initial state scattering models  are unable to reproduce the measured
magnitude of the event--by--event fluctuations in central collisions
of heavy nuclei.
Note that similar conclusion was reached very recently by the analysis
of two pion correlations within the same class of
non--equilibrium models \cite{To:97}.


\section{Event--by--Event Fluctuations \\ in Thermodynamic Models}

At the opposite extreme of the initial state scattering models
without equilibration of the produced particles
are thermodynamic approaches assuming global equilibration,
with the system  possibly undergoing a subsequent collective expansion
until freeze--out.\\
Initial global equilibrium is one of the basic assumptions of the
Landau picture of nuclear collisions \cite{Landau}, where the matter is
fully stopped and globally equilibrated before the onset
of expansion.\\
 In the Bjorken picture \cite{Bjorken} initial termalization
occurs at some proper time $\tau_0$, setting the stage for the subsequent
expansion. The basic difference between the two pictures is that in the
Bjorken one the initial conditions for hydrodynamical expansion are
assumed to be invariant under  Lorentz boost in  longitudinal
direction.\\

Leaving a detailed analysis of the event--by--event fluctuation pattern
in equilibrium models
to future publications, we comment on some points related to the
above discussion. We have seen that the variable $\Phi_{p_T}$ was
designed in such a way that it shows the presence of correlations
between the transverse momentum and multiplicity at the
event--by--event
basis. The starting point was the presence of such correlations in
N+N interactions.
At low collision energies the observed negative correlation can be
interpreted as being due to energy and momentum conservation.
The crucial question is then whether 
thermalization and hydrodynamical expansion can wash out such correlations.

Let us start with the simplest example of an ultrarelativistic gas in
thermal equilibrium. Then the number of particles is determined
by the condition of thermal equilibration itself and the externally defined
volume and total energy of the gas.
This leads to a correlation between entropy (multiplicity) and
temperature (transverse
momentum) at the level of average values ($S \sim T^3$), when the
external conditions vary.
It is important to note,
that the averaging here can be done with respect to a single system by
measuring the quantities in question for a sufficiently long time.
Moreover, the theory
of thermodynamical fluctuations \cite{LL} shows, that the {\it fluctuations}
of entropy and temperature in a given subvolume of a gas
are correlated:
$$
\langle \Delta S \Delta T \rangle = T,
$$
which for massless particles means a positive correlation
between the multiplicity and the transverse momentum.
Therefore, in a thermalized ultrarelativisitc system the particles
are in fact not emitted independently and some definite correlation between
the transverse momentum and multiplicity takes place.
Thus even in the case of full thermalization 
fluctuations due to the non--independent particle emission
in heavy ion collisions should be expected.

The physical origin of the fluctuations
in the  framework of hydrodynamical approach is the same.
Following the above discussion
one may distinguish between two different sources of   fluctuations
in hydrodynamical models.
First of all due to fluctuations in the thermalized energy and volume
occupied by the matter at the early stage initial conditions
vary from event to event.
In Ref. \cite{St:96} it was proposed to describe these early stage
global fluctuations also in terms of thermodynamical fluctuations, where
the role of the heat bath is played by the initial energy of the
colliding nuclei.
These fluctuations are sensitive to the number of effective degrees
of freedom at the early stage  and therefore, if measurable, can serve as
a probe of the properties of the early stage matter.
Finally thermal fluctuations in the elements of freezing--out matter take place
\cite{Sh1}.
Here effective degrees of freedom are hadrons and hadronic resonances.
As suggested in Ref. \cite{Sh1} the magnitude of these fluctuations can
be used to establish freeze--out conditions of hadronic gas.

The exact form  of the finally observed fluctuations
depends on the type of expansion. Despite  the considerable
effort in investigating this problem \cite{Sh, VH, Gyu,
VR} further studies are needed to compare the predictions with the
experimental data.

We conclude,  that in passing from the regime of incoherent
N+N interactions to that of formation of thermalized matter and
its subsequent hydrodynamical expansion, the correlation between the
multiplicity and transverse momentum changes its origin. \\
Thus we can expect changes
in the value of the variable $\Phi_T$.
In the thermalized events showing a collective behaviour one has,
roughly speaking,
 only one 'fluctuating unit', the whole system, whereas in the Cronin
 rescattering case the event is superimposed from a number of sources with
 fluctuating positions in the phase space. A reasonable guess is then that
 the relative "collective" fluctuations measured by $\Phi_T$ should
be smaller than in the Cronin rescattering case and the value of $\Phi_T$
 should thus drop, but further quantitative studies are clearly needed to
 settle this question. Let us also note, that in order to achieve a
 quantitative understanding of event--by--event fluctuations a much better
understanding of the physics of the freeze--out stage is needed (see e.g.
 a recent discussion in Ref. \cite{HS}).

\section{Summary and Final Remarks}

 We have discussed event--by--event transverse momentum fluctuations
in  non--equilibrium and equilibrium models of nucleus--nucleus
collisions in terms of the variable $\Phi_T$.
We have shown that the non--equilibrium initial state scattering
models lead to a large (up to a factor of 10) enhancement
of fluctuations for central Pb+Pb collisions at the CERN SPS in comparison
to that observed in p+p interactions.

The preliminary results of the NA49 Collaboration on $\Phi_T$ fluctuation
measure for central Pb+Pb collisions at 158 A GeV indicate a significant
reduction of these fluctuations relative to
fluctuations in p+p interactions \cite{Rol:97}.
Thus the data show qualitatively different behaviour than that predicted in the
initial state scattering models.
We conclude therefore that such models need to be significantly modified
in order to describe
collisions of heavy nuclei.

We have argued that even in the extreme case of global thermal equilibrium
reached at the early stage of the collision the values of $\Phi_T$
should be larger than zero (independent particle emission limit)
due to the presence of thermal fluctuations at the early and freeze--out stages
and the correlation between the multiplicity (entropy) and transverse momentum
 (temperature) in the equilibrated massless gas.
However as the origin of fluctuations is very different for equilibrium and
non--equilibrium cases a significant change of the $\Phi_{p_T}$ value
is expected to occur
in the course of an increasing equilibration of the created matter.

We hope that the first exciting results of the NA49 Collaboration
will prompt further extensive theoretical and experimental studies.
In particular the measurements of event--by--event fluctuations
in p+p, p+A interactions and central A+A collisions at various collision
energies are necessary.

\section{Acknowledgements}

We would like to thank Mark I. Gorenstein for comments to the manuscript.
The work of A.L. was partially supported by the Russian Foundation for
 Basic Research under Grant 93--02--3815.

\end{document}